\documentclass[aps,pra,twocolumn,showpacs]{revtex4-2}
\usepackage{epsfig,amssymb,amsfonts}
\usepackage{amsmath,amsfonts,amssymb}
\usepackage{graphicx}
\usepackage{caption}
\usepackage{xcolor}
\input{epsf}
\usepackage{epstopdf}
\usepackage{natbib}

\usepackage[labelformat=simple]{subcaption}
\makeatletter
\renewcommand\p@subfigure{\thefigure}
\makeatother

\begin{document}

\title{Optimal STIRAP shortcuts using the spin to spring mapping}
\author{Vasileios Evangelakos}
\author{Emmanuel Paspalakis}
\author{Dionisis Stefanatos}
\email{dionisis@post.harvard.edu}
\affiliation{Materials Science Department, School of Natural Sciences, University of Patras, Patras 26504, Greece}
\date{\today}
\begin{abstract}
We derive shortcuts to adiabaticity maximizing population transfer in a three-level $\Lambda$ quantum system, using the spin to spring mapping to formulate the corresponding optimal control problem on the simpler system of a classical driven dissipative harmonic oscillator. We solve the spring optimal control problem and obtain analytical expressions for the impulses, the durations of the zero control intervals and the singular control, which are the elements composing the optimal pulse sequence. We also derive suboptimal solutions for the spring problem, one with less impulses than the optimal and others with smoother polynomial controls. We then apply the solutions derived for the spring system to the original system, and compare the population transfer efficiency with that obtained for the original system using numerical optimal control. For all dissipation rates used, the efficiency of the optimal spring control approaches that of the numerical optimal solution for longer durations, with the approach accomplished earlier for smaller decay rates. The efficiency achieved with the suboptimal spring control with less impulses is very close to that of the optimal spring control in all cases, while that obtained with polynomial controls lies below, and this is the price paid for not using impulses, which can quickly build a nonzero population in the intermediate state. The analysis of the optimal solution for the classical driven dissipative oscillator is not restricted to the system at hand but can also be applied in the transport of a coherent state trapped in a moving harmonic potential and the transport of a mesoscopic object in stochastic thermodynamics.
\end{abstract}

\maketitle

\section{Introduction}

Stimulated Raman adiabatic passage (STIRAP) is a prominent pulse sequence for transferring robustly the population between two indirectly linked states in a three-level quantum system \cite{Bergmann98,Kral07,Vitanov17,Sola18,Bergmann19,Vitanov97,Kobrak98}. The population is transferred from the starting state $|1\rangle$ to the final state $|3\rangle$ throughout a mid level $|2\rangle$, which is subject to dissipation, connected to the other levels using the pump and Stokes laser fields. The signature of STIRAP pulse sequence is the counterintuitive order of the Stokes and pump pulses, resulting in the formation of a coherent superposition between states $|1\rangle$ and $|3\rangle$ which, in the course of time, evolves from the initial state $|1\rangle$ to the target state $|3\rangle$. If the Stokes and pump laser fields are gradually (adiabatically) modified, the lossy intermediate state $|2\rangle$ is scarcely inhabited, attaining thus a high fidelity population transfer. The significant advantage of STIRAP is the resilience against moderate modifications of the experimental parameters. This is the cause behind its popularity in several areas of current physics \cite{Bergmann19}, like optical waveguides \cite{Paspalakis06,Dreisow09}, nitrogen-vacancy centers in diamond \cite{Golter14}, matter waves \cite{Menchon16}, atomic systems \cite{Liedl22,Schutz22,Stabel22}, superconducting quantum circuits \cite{Kumar16,Dogra22b}, quantum dots coupled to plasmonic nanostructures \cite{Domenikou22}, and quantum information processing \cite{Genov23}.

Optimal control \cite{Bryson75} has been used throughout the years in order to improve STIRAP fidelity, as in several other important quantum systems \cite{Glaser15,StefanatosReview,Boscain21}. Despite that the counterintuitive sequence with zero initial values of the pulses was proved not to be optimal for maximizing the target population \cite{Band94}, it was later demonstrated numerically that it turns to the optimal solution if the cost function is augmented by a term penalizing the intermediate state occupation \cite{Sola99,Kis02,Kumar11}. In another study \cite{Rat12}, where the methods developed in Refs. \cite{Khaneja03,Stefanatos04,Stefanatos05} were used with bounded Stokes and unbounded pump fields, an optimal solution imitating the counterintuitive sequence was derived, with a constant Stokes pulse and a pump pulse smaller at the beginning and larger at the end. In Ref. \cite{Assemat12}, where a cost function punishing the middle state habitation was used, the optimal solution turned out to be a counterintuitive sequence with nonzero Stokes and pump pulses at the initial and final times, respectively. When dissipation is omitted and the objective is set to minimize the transfer time from $|1\rangle$ to $|3\rangle$ or the fluence of the pulses, the intuitive pulse sequence (pump field before Stokes) arises as the optimal solution \cite{Boscain02}. Note that the minimum time transfer problem in the absence of dissipation was also investigated in Ref. \cite{Dalessandro20}.
Along with the above investigations, which utilize analytical and numerical optimal control methods to increase STIRAP transfer fidelity, several variants of shortcuts to adiabaticity \cite{STA19} have also been exploited \cite{Demirplak05,Vasilev09,Chen10b,Giannelli14,Masuda15,Li16,Clerk16,Santos17,Santos20,Kolbl19,Dridi20,Petiziol20,Blekos20,Zheng22,Laforgue22,Dogra22}.

In our recent study \cite{Stefanatos21a}, we solved the problem of maximizing population transfer in a $\Lambda$-type system for the special case where dissipation is much larger than the amplitude of the applied fields, so the intermediate state $|2\rangle$ can be adiabatically eliminated from the equations, leading thus to a simpler two-dimensional problem. In a subsequent work \cite{Stefanatos22royal}, we solved the full three-dimensional optimal control problem, which can be formulated as a problem on a spin-$1/2$ thus we call it in the following the \emph{spin} problem, elucidating the role of singular control in the optimal solution. Specifically, we explained that it is this singular control which takes advantage of the longer available durations to identify optimal paths along which the effect of dissipation is minimized, achieving thus higher transfer fidelity. In that work, although some analytical steps have been performed, for example the singular control was expressed in terms of a feedback law, we eventually have recourse to numerical optimal control \cite{bocop} for finding the optimal pulse sequence.

In the present study we use the spin to spring mapping \cite{Li17,Martikyan20b} to formulate the corresponding optimal control problem on the simpler system of a classical driven dissipative harmonic oscillator. We solve this spring optimal control problem finding analytical expressions for the impulses, the durations of the zero control intervals, and the singular control. We also derive suboptimal solutions for the spring problem, one with less impulses than the optimal and others with smoother polynomial controls. We then apply the solutions derived for the spring system to the original \emph{spin} system, and compare the population transfer efficiency with that obtained for the original system using numerical optimal control \cite{Stefanatos22royal}. We find that for all dissipation rates used, the efficiency of the optimal spring control approaches that of the numerical optimal spin control for longer durations, with the approach accomplished earlier for smaller decay rates. The efficiency achieved with the suboptimal spring control with less impulses is very close to that of the optimal spring control in all cases, while that obtained with polynomial controls lies below, and this is the price paid for not using impulses, which can quickly build a nonzero population in the intermediate state. Note that the present analysis of the optimal solution for the classical driven dissipative oscillator is not restricted only to the system at hand but can also find applications in the transport of a coherent state trapped in a moving harmonic potential and the transport of a mesoscopic object in stochastic thermodynamics \cite{Zhang22}.

The article has the following structure. In Sec. \ref{sec:system} we formulate the spin optimal control problem, while in section \ref{sec:mapping} we formulate and analyze the corresponding spring optimal control problem. In Sec. \ref{sec:idealcase} we present optimal and suboptimal solutions to the spring problem, which are applied in Sec. \ref{sec:results} to the original spin problem. The findings of the present study are summarized in section \ref{sec:conclusion}.

\section{The spin optimal control problem}

\label{sec:system}

If
\begin{equation}
\label{psi}
|\psi(t)\rangle= c_1(t)|1\rangle+c_2(t)|2\rangle+c_3(t)|3\rangle
\end{equation}
denotes the state of the $\Lambda$-system then, under one-photon and two-photon resonance, the probability amplitudes evolve according to the Schr\"{o}dinger equation
\begin{equation}
\label{Schrodinger}
i
\left(\begin{array}{c}
    \dot{c}_1  \\
    \dot{c}_2 \\
    \dot{c}_3
\end{array}\right)
=
\frac{1}{2}
\left(\begin{array}{ccc}
    0 & \Omega_p(t) & 0 \\
    \Omega_p(t) & -i\Gamma & \Omega_s(t) \\
    0 & \Omega_s(t) & 0
  \end{array}\right)
\left(\begin{array}{c}
    c_1  \\
    c_2 \\
    c_3
\end{array}\right),
\end{equation}
where $\Omega_p(t), \Omega_s(t)$ denote the Rabi frequencies of the pump and Stokes fields, respectively, and $\Gamma$ the dissipation rate from the excited state $|2\rangle$. At $t=0$ the whole population is in state $|1\rangle$
\begin{equation}
\label{initial_conditions}
c_1(0)=1, \quad c_2(0)=c_3(0)=0,
\end{equation}
and the goal is to determine the control fields $\Omega_p(t), \Omega_s(t)$ which maximize the population $|c_3(T)|^2$ transferred to state $|3\rangle$ up to some given final time $t=T$.

To make the above optimization problem meaningful, it is necessary to impose a bound on the control amplitudes. This is accomplished through the constraint
\begin{equation}
\label{bounded_fluence}
\Omega^2_p(t)+\Omega^2_s(t)=\Omega^2_0\quad\mbox{constant},
\end{equation}
fixing the amplitude of the total field in the equivalent two-level picture to the constant value $\Omega_0$ and leaving as the only control variable its direction, determined from the
mixing angle
\begin{equation}
\label{mixing_angle}
\tan{\theta(t)}=\frac{\Omega_p(t)}{\Omega_s(t)}.
\end{equation}
By using the following state variables \cite{Boscain02,Vitanov06}
\begin{equation}
\label{transform1}
Z=c_1,\quad Y=-ic_2,\quad X=-c_3,
\end{equation}
and normalized time $\Omega_0t$,
we get the equivalent system
\begin{equation}
\label{XYZ}
\left(\begin{array}{c}
    \dot{Z}  \\
    \dot{Y} \\
    \dot{X}
\end{array}\right)
=
\frac{1}{2}
\left(\begin{array}{ccc}
    0 & \sin\theta(t) & 0 \\
    -\sin\theta(t) & -\Gamma & \cos\theta(t) \\
    0 & -\cos\theta(t) & 0
  \end{array}\right)
\left(\begin{array}{c}
    Z  \\
    Y \\
    X
\end{array}\right),
\end{equation}
free from complex quantities. Note that Eq. (\ref{XYZ}) describes the evolution of a spin with Bloch vector $(X, Y, Z)^T$ under the influence of a total field $(\sin{\theta}, 0, \cos{\theta})^T/2$, while there is also dissipation $-\Gamma/2$ in coordinate $Y$. Since $Z^2=|c_1|^2, Y^2=|c_2|^2, X^2=|c_3|^2$, the Bloch vector starts at $t=0$ from the north pole and the target at $t=T$ is the maximization of $X^2(T)=|c_3(T)|^2$. Moreover, note that when $\theta=0$ the total field is aligned to the north pole and coincides with the Stokes field, while for $\theta=\pi/2$ is aligned to the $X$-axis and coincides with the pump field.

Instead of working in the above $XYZ$ reference frame, we will use the transformation \cite{Ivanov05}
\begin{eqnarray}
z&=&Z\cos\theta+X\sin\theta,\label{t_z}\\
y&=&Y,\label{t_y}\\
x&=&Z\sin\theta-X\cos\theta,\label{t_x}
\end{eqnarray}
where the new coordinates $z, x$ represent the dark and bright states of the quantum system, respectively, and $y$ still corresponds to the intermediate state $|2\rangle$. The transformed $xyz$ system becomes
\begin{equation}
\label{xyz}
\left(\begin{array}{c}
    \dot{z}  \\
    \dot{y} \\
    \dot{x}
\end{array}\right)
=
\left(\begin{array}{ccc}
    0 & 0 & -\omega(t) \\\noalign{\vskip3pt}
    0 & -\frac{\Gamma}{2} & -\frac{1}{2} \\\noalign{\vskip3pt}
    \omega(t) & \frac{1}{2} & 0
  \end{array}\right)
\left(\begin{array}{c}
    z  \\
    y \\
    x
\end{array}\right),
\end{equation}
and observe that now the derivative of the mixing angle $\omega(t)$ emerges as the control variable,
\begin{equation}
\label{theta}
\dot{\theta}=\omega(t).
\end{equation}
The boundary conditions
\begin{equation}
\label{b_theta}
\theta(0)=0,\quad\theta(T)=\frac{\pi}{2},
\end{equation}
are applied on the mixing angle, so in the original $XYZ$ frame the total field is aligned initially with the north pole (as well as the initial Bloch vector) and finally with the $X$-axis, hence imitating shortcuts to adiabaticity. If we use in Eqs. (\ref{t_z})-(\ref{t_x}) the initial conditions for $X, Y, Z$ as well as the boundary conditions for $\theta$, we obtain the initial conditions for the transformed coordinates $x,y,z$
\begin{equation}
\label{starting point}
x(0)=y(0)=0, \quad z(0)=1,
\end{equation}
and moreover the quantity to be maximized, $z(T)=X(T)$ at the given final time $t=T$. In our previous work \cite{Stefanatos22royal} we solved this optimal control under the constraint $\omega(t)\geq 0$,
which assures that $\theta$ is monotonically increasing. In this work we will address the same problem by taking advantage of the spins to springs mapping in order to make easier the derivation of the optimal control, which we will then test to the original spin system. Note that if we do not place the above restriction on $\theta$, then we know from our recent work \cite{Stefanatos22royal} as well as Ref. \cite{Boscain02} that in the absence of dissipation the intuitive pulse sequence (with the pump pulse preceding the Stokes pulse) arises as the optimal solution. We thus use this condition in order to avoid such situations and preserve the counterintuitive pulse order which is the signature of STIRAP and is also associated with increased robustness.

\section{Spin to spring mapping and analysis of the spring optimal control problem}

\label{sec:mapping}

The transformed frame $xyz$ is appropriate for exploiting the mapping from spins to springs, to facilitate the optimal control design \cite{Li17,Martikyan20b}. The reason is that this mapping is effective for small polar angles of the Bloch vector \cite{Martikyan20b}, something which is expected to be the case for the transformed frame $xyz$, since $z(0)=1$ initially and the goal is to maximize the final $z(T)$, especially for longer durations $T$. The spring system corresponding to the spin system (\ref{xyz}) can be easily obtained by defining the new control $u(t)$ as
\begin{equation}
\label{spintospringset}
u(t)=\omega(t)z(t)=\dot \theta(t)z(t),
\end{equation}
which leads to
\begin{equation}
\label{spring_system}
\left(\begin{array}{c} \dot y\\ \dot x\end{array}\right)=
\left(\begin{array}{cc} -\frac{\Gamma}{2} & -\frac{1}{2}\\ \frac{1}{2} & 0\end{array}\right)
\left(\begin{array}{c} y\\ x\end{array}\right)
+u\left(\begin{array}{c} 0\\ 1\end{array}\right).
\end{equation}
By taking the derivative of $\dot{y}$ in Eq. (\ref{spring_system}) and substituting the expression for $\dot{x}$, we obtain the second order differential equation
\begin{equation}
\label{osc_equ}
\ddot y=-\frac{\Gamma}{2} \dot y -\frac{1}{4} y-\frac{1}{2}u,
\end{equation}
which obviously describes the motion of a driven-damped harmonic oscillator with damping coefficient $\Gamma/2$, angular frequency $\omega_0=1/2$, and external force $u/2$.

Instead of using the state variable $x$, we will use the ``velocity"
\begin{equation}
\label{v}
v=\dot{y}=-\frac{1}{2}(\Gamma y+x),
\end{equation}
which will allow us to intuitively understand the optimal solutions. The spring system (\ref{spring_system}) then becomes
\begin{equation}
\label{spring_1}
\left(\begin{array}{c} \dot y\\ \dot v\end{array}\right)=
\left(\begin{array}{cc} 0 & 1\\ -\frac{1}{4} & -\frac{\Gamma}{2}\end{array}\right)
\left(\begin{array}{c} y\\ v\end{array}\right)
+u\left(\begin{array}{c} 0\\ -\frac{1}{2}\end{array}\right).
\end{equation}
The initial conditions (\ref{starting point}) for $x, y$ are translated to
\begin{equation}
\label{start}
y(0)=v(0)=0
\end{equation}
corresponding to an oscillator starting from equilibrium. In the transformed frame $xyz$ the goal is to maximize the terminal value $z(T)$. Since the state variable $z$ of the spin system is absent from the corresponding spring system, in order to indirectly accomplish this target, we require the minimization of the following cost
\begin{equation}
\label{cost}
J=\Gamma\int_0^Ty^2(t)dt,
\end{equation}
accompanied with the terminal conditions $x(T)=y(T)=0$. Note that, since for $\boldsymbol{r}=(x, y, z)^T$ it is $dr^2/dt^2=-\Gamma y^2$ from the spin system (\ref{xyz}), the cost $J$ to be minimized expresses the losses in the Bloch vector due to dissipation, and under the imposed terminal conditions it is $r(T)=z(T)$. These terminal conditions are translated to
\begin{equation}
\label{end}
y(T)=v(T)=0,
\end{equation}
thus the spring should also return back to equilibrium. For the control $u(t)$ we will use the condition
\begin{equation}
\label{impulse}
\int_{0}^{T} u(t) \,dt=\frac{\pi}{2}.
\end{equation}
Note that this condition is consistent with Eqs. (\ref{b_theta}), (\ref{spintospringset}) only when $z(t)$ remains close to unity throughout. In order to resolve this issue and assure that $\theta(T)=\pi/2$ so the total field is aligned with $X$-axis in the original $XYZ$ frame, after solving the spring optimal control problem we will use
\begin{equation}
\label{dot_theta}
\dot{\theta}=u(t).
\end{equation}
This approach may lead to some small remainders in $Y(T)$ and $Z(T)$ which are wiped out by the dissipative term $-\Gamma Y/2$ in Eq. (\ref{XYZ}) if the evolution is extended beyond $T$, while the equation for $X$ is decoupled for the final $\theta=\pi/2$.

Now we can state the optimal control problem for the spring system (\ref{spring_1}) as follows. We would like to find the force, proportional to the control $u(t)$, with impulse fixed by the integral condition (\ref{impulse}), which within a given duration $T$ returns the system to its initial equilibrium position, while minimizes a cost proportional to the mean squared displacement $\langle y^2\rangle=\int_0^Ty^2(t)dt/T$. We also require the control (and thus the force) to be nonnegative
\begin{equation}
\label{pos_u}
u(t)\geq 0,
\end{equation}
so the angle $\theta$ is monotonically increasing.

For the solution of the above optimal control problem we form the corresponding
control Hamiltonian \cite{Bryson75}, a
mathematical construction whose minimization leads to the minimization of cost $J$. It is formulated by adjoining to each state equation, to the cost $J$ and the control constraint (\ref{impulse}) a Lagrange multiplier:
\begin{eqnarray}
\label{Hc}
H_c&=&\lambda_yv+\lambda_v\left(-\frac{1}{4}y-\frac{\Gamma}{2}v-\frac{1}{2}u\right)+\mu u+\kappa\Gamma y^2 \\
   &=& \left(\mu-\frac{1}{2}\lambda_v\right)u+\lambda_yv-\frac{1}{4}\lambda_vy-\frac{\Gamma}{2}\lambda_vv+\kappa\Gamma y^2. \nonumber
\end{eqnarray}
Here, $\lambda_y, \lambda_v$ are Lagrange multipliers corresponding to state variables $y, v$ and satisfy the adjoint equations
\begin{eqnarray}
\dot{\lambda}_y&=&-\frac{\partial H_c}{\partial y}=\frac{1}{4}\lambda_v-2\kappa\Gamma y, \label{ly}\\
\dot{\lambda}_v&=&-\frac{\partial H_c}{\partial v}=-\lambda_y+\frac{\Gamma}{2}\lambda_v, \label{lv}
\end{eqnarray}
while $\mu, \kappa $ are constant multipliers corresponding to the cyclic ``variables" $J$ and $\theta$ through Eqs. (\ref{cost}) and (\ref{impulse}), respectively.

As optimal control theory \cite{Bryson75} dictates, the selection of the control function $u(t)$ is made such that the control Hamiltonian $H_c$ is minimized. In the present case where the control is unbounded from above, see Eq. (\ref{pos_u}), even infinite values are allowed momentarily, corresponding to finite negative instantaneous impulses in velocity and positive instantaneous jumps in the angle $\theta$, as implied by Eqs. (\ref{spring_1}) and (\ref{dot_theta}), respectively. Considering that $H_c$ is a linear function of $u$ with coefficient the \emph{switching} function
\begin{equation}
\label{switching_function}
\phi=\mu-\frac{1}{2}\lambda_v,
\end{equation}
if $\phi<0$ for a finite time interval then the corresponding optimal control should be infinite for the whole interval, which is clearly nonphysical. We deduce that in the case of $u$ unbounded from above, the condition $\phi\geq0$ should hold almost everywhere, excluding some isolated points where jumps in $v$ and $\theta$ may take place. For time intervals where $\phi>0$ the corresponding optimal control is $u(t)=0$, equal to the lower bound in Eq. (\ref{pos_u}), in order to minimize $H_c$. There might also exist time intervals where $\phi=0$. In such cases the optimal control cannot be deduced from the minimization of $H_c$ and is named \emph{singular} \cite{Bryson75}. Singular controls often constitute essential parts of the optimal pulse sequence, as for example in the famous aerospace Goddard problem \cite{Tsiotras92}, and have been also exploited in dissipative two-level systems for minimizing losses \cite{Lapert10,Lin20}. The singular control is determined from the condition $\phi=0$ holding on the singular arc. From $\phi=0$ and Eq. (\ref{switching_function}) we get $\lambda_v=2\mu$ constant. Consequently $\dot{\lambda_v}=0$, and from Eq. (\ref{lv}) we find that $\lambda_y=\Gamma\lambda_v/2$, also a constant. Thus $\dot{\lambda_y}=0$, and from Eq. (\ref{ly}) we obtain $y=\lambda_v/(8\kappa\Gamma)=\mu/(4\kappa\Gamma)$. This means that on a singular arc the position $y$ is held constant, $y=y_s$, which also implies a zero velocity $v_s=0$, while from the spring equation (\ref{spring_1}) the singular control is found to be
\begin{equation}
\label{us}
u_s=-\frac{1}{2}y_s,
\end{equation}
balancing the spring restoring force.

\section{Solutions to the spring problem}

\label{sec:idealcase}

From the previous analysis becomes obvious that the optimal control may contain positive instantaneous impulses and finite intervals where it is zero (off), called bang controls since it assumes values on the boundaries of the allowed range (\ref{pos_u}), as well as finite intervals where it takes constant positive values, intermediate between zero and infinity, which are called singular controls. In the following we evaluate the performance of pulse sequences composed of such pulse elements. We also consider smoother polynomial pulses.

\subsection{An intuitive sub-optimal pulse sequence}

We begin by considering a bang-bang-singular-bang-bang pulse sequence composed by impulse-off-singular-off-impulse pulses, mathematically expressed as
\begin{equation}
\label{almost_optimal_control}
u(t)=v_1\delta(t)+u_c\Theta(t-t_1)\Theta(t_2-t)+v_2\delta(t-T),
\end{equation}
where $\delta(t)$ and $\Theta(t)$ are the Dirac and Heaviside functions, respectively. An impulse (kick) at $t=0^+$ gives instantaneously an initial velocity to the system, which is then left to evolve freely until time $t=t_1$, where its velocity becomes momentarily zero. At this point a constant force (singular control) is applied to``freeze" it there until time $t=t_2$, when the oscillator is released and left to evolve freely again for the rest of the duration $T-t_2$. As the system passes through the equilibrium position $y=0$ at time $t=T^-$, a final impulse is applied in order to stop the oscillator there. We move to find analytical expressions for the initial and final impulses $v_1, v_2$, the switching times $t_1,t_2$ and the singular control $u_s$, as functions of the given duration $T$ and the decay rate $\Gamma$. In order to achieve this we break the evolution down to three parts. 

After the first impulse, the system acquires a initial velocity $v(0^+)=\dot y(0^+)=-v_1/2$, while it remains at $y(0^+)=0$. During the interval $0<t<t_1$ the control is turned off, $u(t)=0$. From Eq. (\ref{osc_equ}) and for the given initial conditions we obtain the following solution in this interval
\begin{equation}
\label{y1_almopt}
    y(t)=-\frac{2v_1 e^{-\frac{\Gamma}{4} t} \sin \left(\frac{1}{4} \sqrt{4-\Gamma ^2} t\right)}{\sqrt{4-\Gamma ^2}},
\end{equation}
where note that here we concentrate on cases with $\Gamma<2$; for $\Gamma\geq 2$ the approach described in Ref. \cite{Stefanatos21a} can be followed.
The switching time $t_1$ is obtained form the requirement that the velocity becomes momentarily zero, $v(t_1)=\dot y(t_1)=0$. Using Eq. (\ref{y1_almopt}) we find
\begin{equation}
\label{t1_almopt}
    t_1=\frac{4 \tan ^{-1}\left(\frac{\sqrt{4-\Gamma ^2}}{\Gamma }\right)}{\sqrt{4-\Gamma ^2}}.
\end{equation}
During the interval $t_1<t<t_2$ a constant singular control $u(t)=u_s$ is applied, balancing the restoring spring force and maintaining the system at the constant position $y_s=y(t_1)$. From Eq. (\ref{us}) we find
\begin{equation}
\label{u1_almopt}
    u_s=-\frac{1}{2} y_s=-\frac{1}{2} y(t_1)=\frac{1}{2}v_1 e^{-\frac{\Gamma  \tan ^{-1}\left(\frac{\sqrt{4-\Gamma ^2}}{\Gamma }\right)}{\sqrt{4-\Gamma ^2}}},
\end{equation}
where note that $u_s>0$ for $v_1>0$.
At $t=t_2$ the system is released and left to evolve freely in the interval $t_2<t<T$. For $u(t)=0$ and initial conditions $y(t_2)=y(t_1)$, $v(t_2)=\dot y_3(t_2)=0$, we find the solution
\begin{equation}
\begin{split}
y(t)&=-2u_s e^{-\frac{\Gamma}{4} (t-t_2)} \left\{ \cos \left[\frac{1}{4} \sqrt{4-\Gamma ^2} (t-t_2)\right] \right. \\
&\left.+\frac{\Gamma}{\sqrt{4-\Gamma ^2}}\sin \left[\frac{1}{4} \sqrt{4-\Gamma ^2} (t-t_2)\right] \right\}
\end{split}
\end{equation}
The requirement $y(T)=0$ gives
\begin{equation}
\label{t2_almopt}
    t_2=T-\frac{4\left[ \pi -\tan ^{-1}\left(\frac{\sqrt{4-\Gamma ^2}}{\Gamma }\right)\right]}{\sqrt{4-\Gamma ^2}}.
\end{equation}
Note that the ordering $t_2>t_1$, using Eqs. (\ref{t1_almopt}) and (\ref{t2_almopt}), leads to the condition
\begin{equation}
\label{minT}
   T>\frac{4 \pi }{\sqrt{4-\Gamma ^2}}
\end{equation}
which is the minimal duration needed for a singular arc to appear in the trajectory. For smaller durations the pulse sequence under consideration has the bang-bang form impulse-off-impulse.
The final impulse is such that the system stops at $y(T)=0$, thus $v(T)=v(T^-)-v_2/2=0$ and
\begin{equation}
\label{v2_almopti}
    v_2=2v(T^-)=2\dot y(T^-)=2 u_s e^{\frac{\Gamma\left[\tan ^{-1}\left(\frac{\sqrt{4-\Gamma ^2}}{\Gamma }\right)-\pi\right]}{\sqrt{4-\Gamma ^2}}}.
\end{equation}
By substituting Eq.(\ref{u1_almopt}) we have
\begin{equation}
\label{v2_almopt}
    v_2=v_1 e^{-\frac{\pi  \Gamma }{\sqrt{4-\Gamma ^2}}},
\end{equation}
thus the final impulse to stop the motion is smaller than the first one that initiated it, as expected for a dissipative system. Now observe that, if we use expression (\ref{almost_optimal_control}) for the control in condition (\ref{impulse}), then the latter becomes
\begin{equation}
\label{th(T)}
    v_1+(t_2-t_1)u_s+v_2=\frac{\pi}{2}.
\end{equation}
By substituting Eqs. (\ref{u1_almopt}) and (\ref{v2_almopt}) in Eq. (\ref{th(T)}) we can solve for $v_1$ and derive analytical expressions for $v_1, u_s, v_2$ in terms of $\Gamma, T$, which can later be used in the control of the original system.


\subsection{Optimal control}

In the derivation of the suboptimal pulse sequence we used the intuitive condition that the velocity is zero before entering and after exiting the singular arc. But the actual optimal solution is obtained from the requirement that the switching function $\phi$, given in Eg. (\ref{switching_function}), and not the velocity, is zero at the switching times $t_1, t_2$. It is thus possible the oscillator to have nonzero velocity before entering and after exiting the singular arc and, after some thought, this observation can lead us to the optimal pulse sequence. As in the previous case, an impulse $v_1$ is applied at $t=0^+$ and the system is again left to evolve freely, but now until a time $t=t_1$ beyond the turning point where the velocity becomes zero, thus the oscillator has started to move back to $y=0$. At $t=t_1^-$ a second impulse $v_2$ is applied annihilating the oscillator's velocity so the system can enter the singular arc, which is enforced with the application of a constant singular control $u(t)=u_s$, freezing the dynamics until time $t=t_2$. The advantage of this strategy compared to the suboptimal one is that, with the appropriate choice of the parameters, the constant $y(t)=y_s$ on the singular arc can be made closer to zero, leading thus to a smaller cost $J$ in Eq. (\ref{cost}). This observation will become obvious below, when considering specific examples. Immediately after the release of the singular force, at $t=t_2^+$, a third impulse $v_3$ is applied such that the oscillator reaches the equilibrium position $y=0$ at time $t=T^-$. There, a final impulse is used to stop the oscillator. The optimal control can be expressed in compact form as
\begin{equation}
\begin{split}
\label{optimal_control}
v(t)&=v_1\delta(t)+u_s\Theta(t-t_1)\Theta(t_2-t) \\
&+v_2\delta(t-t_1)+v_3\delta(t-t_2)+v_4\delta(t-T)
\end{split}
\end{equation}
In order to find the impulses $v_1, v_2, v_3, v_4$, the constant singular force $u_s$ and the switching times $t_1, t_2$, we break the dynamics down to three stages, as before. We are also going to use the rescaled time
\begin{equation}
\tilde t=\frac{\sqrt{4-\Gamma ^2}}{4}t
\end{equation}
in order to simplify the expressions.

After the first impulse and in the interval $0<t<t_1$ the solution (\ref{y1_almopt}) is still valid, and in terms of the rescaled time is expressed as
\begin{equation}
\label{y1_opt}
    y(t)=-\frac{2 v_1 e^{-\frac{\Gamma  \tilde{t}}{\sqrt{4-\Gamma ^2}}} \sin \left(\tilde{t}\right)}{\sqrt{4-\Gamma ^2}}
\end{equation}
At $t=t_1^-$ the second impulse $v_2$ annihilates the acquired velocity, thus $v(t_1)=v(t_1^-)-v_2/2=\dot{y}(t_1^-)-v_2/2=0$ and
\begin{equation}
\label{v2_opt}
    v_2=v_1 e^{-\frac{\Gamma  \tilde{t}_1}{\sqrt{4-\Gamma ^2}}} \left[\frac{\Gamma  \sin \left(\tilde{t}_1\right)}{\sqrt{4-\Gamma ^2}}-\cos \left(\tilde{t}_1\right)\right].
\end{equation}
In the interval $t_1<t<t_2$ the constant singular control $u(t)=u_s$ freezes the dynamics at the constant position $y_s=y(t_1)$, thus
\begin{equation}
\label{u1_opt}
    u_s=-\frac{1}{2} y_s=-\frac{1}{2} y(t_1)=v_1 e^{-\frac{\Gamma  \tilde{t}_1}{\sqrt{4-\Gamma ^2}}}\frac{\sin \left(\tilde{t}_1\right)}{\sqrt{4-\Gamma ^2}}.
\end{equation}
The system is released and a third impulse $v_3$ is immediately applied at $t=t_2^+$. For $u(t)=0$ and initial conditions $y(t_2)=y(t_1)$, $v(t_2^+)=v(t_2)-v_3/2=-v_3/2$, we find
in the interval $t_2<t<T$ the solution
\begin{equation}
\label{y3_opt}
\begin{split}
    y(t) &= \frac{-2 e^{\frac{\Gamma  \left(\tilde{t}_2-\tilde{t}\right)}{\sqrt{4-\Gamma ^2}}}}{\sqrt{4-\Gamma ^2}} \times \Big\{ v_3 \sin \left(\tilde{t}-\tilde{t}_2\right) \\
   & +u_s \left[\sqrt{4-\Gamma ^2} \cos \left(\tilde{t}-\tilde{t}_2\right)+\Gamma  \sin \left(\tilde{t}-\tilde{t}_2\right)\right] \Big\}.
\end{split}
\end{equation}
The final time requirement $y(T)=0$ leads to the relation
\begin{equation}
\label{v3_opt}
    v_3=-u_s \left[\sqrt{4-\Gamma ^2} \cot \left(\tilde{T}-\tilde{t}_2\right)+\Gamma \right].
\end{equation}
Additionally, a final impulse $v_4$ is applied at $t=T$ in order to stop the oscillator motion. The corresponding condition $v(T)=\dot y(T^-)-v_4/2=0$ leads to the expression
\begin{equation}
\label{v4_opt}
    v_4=\sqrt{4-\Gamma ^2} u_s e^{\frac{\Gamma  \left(\tilde{t}_2-\tilde{T}\right)}{\sqrt{4-\Gamma ^2}}} \csc \left(\tilde{T}-\tilde{t}_2\right).
\end{equation}
We can now express parameters $v_2, v_3, v_4, u_s$ as functions of the initial impulse $v_1$ and the switching times $t_1, t_2$ as
\begin{equation}
\label{opt_par_all}
\begin{array}{ccc}
v_2&=&v_1 e^{-\frac{\Gamma  \tilde{t}_1}{\sqrt{4-\Gamma ^2}}} \left[\frac{\Gamma  \sin \left(\tilde{t}_1\right)}{\sqrt{4-\Gamma ^2}}-\cos \left(\tilde{t}_1\right)\right]\\
v_3&=&-\frac{v_1 e^{-\frac{\Gamma  \tilde{t}_1}{\sqrt{4-\Gamma ^2}}} \sin \left(\tilde{t}_1\right) \left[\sqrt{4-\Gamma ^2} \cot \left(\tilde{T}-\tilde{t}_2\right)+\Gamma \right]}{\sqrt{4-\Gamma ^2}}\\
v_4&=&v_1 \sin \left(\tilde{t}_1\right) e^{-\frac{\Gamma  \left(\tilde{t}_1-\tilde{t}_2+\tilde{T}\right)}{\sqrt{4-\Gamma ^2}}} \csc \left(\tilde{T}-\tilde{t}_2\right)\\
u_s&=&\frac{v_1 e^{-\frac{\Gamma  \tilde{t}_1}{\sqrt{4-\Gamma ^2}}} \sin \left(\tilde{t}_1\right)}{\sqrt{4-\Gamma ^2}}
\end{array}
\end{equation}
For the optimal control (\ref{optimal_control}), condition (\ref{impulse}) gives
\begin{equation}
\label{th(T)_opt}
    v_1+v_2+(t_2-t_1)u_c+v_3+v_4=\frac{\pi}{2}.
\end{equation}
By substituting Eqs. (\ref{opt_par_all}) we can express $v_1$ in terms of $t_1, t_2$.

The switching times  $t_1, t_2$ will be obtained from the nullification of the switching function using the adjoint system (\ref{ly}), (\ref{lv}). We will actually exploit the ``normalized" Lagrange multipliers $\tilde{\lambda}_y=\lambda_y/\mu$, $\tilde{\lambda}_v=\lambda_v/\mu$, satisfying the equations
\begin{eqnarray}
\dot{\tilde{\lambda}}_y&=&\frac{1}{4}\tilde{\lambda}_v-\frac{1}{2}\frac{y}{y_s}, \label{ly_t}\\
\dot{\tilde{\lambda}}_v&=&-\tilde{\lambda}_y+\frac{\Gamma}{2}\tilde{\lambda}_v, \label{lv_t}
\end{eqnarray}
where we have used the relation $y_s=\mu/(4\kappa\Gamma)$ for the constant position on the singular arc to replace the ratio $\kappa/\mu$ of the constant multipliers, see the discussion at the end of Sec. \ref{sec:mapping}.

In the interval $0\leq t\leq t_1$, $y(t)$ is given by expression (\ref{y1_opt}) while $y_s=y(t_1)$. We observe that the ratio $y/y_s$ appearing in Eq. (\ref{ly_t}) is independent of $v_1$ and depends only on time $t_1$ through $y_s$. The switching function (\ref{switching_function}) is expressed as $\phi=\mu (1-\tilde{\lambda}_v/2)$. Since $\mu\neq 0$, when the switching function is zero then $\tilde{\lambda}_v=2$. On the singular arc, the multipliers have the constant values $\tilde{\lambda}_v=2$ and $\tilde{\lambda}_y=\Gamma\tilde{\lambda}_v/2=\Gamma$. Since the impulses do not affect the oscillator position $y$ but only its velocity $v$, the multipliers are also unaffected. Consequently, the multipliers are not affected by the second impulse but assume the values which they have on the singular arc, $\tilde{\lambda}_v(t_1)=2$ and $\tilde{\lambda}_y(t_1)=\Gamma$. Using these terminal conditions, we can propagate backwards Eqs. (\ref{ly_t}), (\ref{lv_t}), from $t=t_1$ to $t=0$ immediately after the first impulse, and obtain
\begin{equation}
\label{lv(0)}
\tilde{\lambda}_v(0)=A_1 \cosh\left(\frac{t_1 \Gamma}{4}\right)+B_1 \sinh\left(\frac{t_1 \Gamma}{4}\right),
\end{equation}
where
\begin{equation}
\label{A1_B1}
\begin{array}{ccc}
A_1&=&\frac{\Gamma  \sin \left(\tilde{t}_1\right)}{\sqrt{4-\Gamma ^2}}+\cos \left(\tilde{t}_1\right),\\[7pt]
B_1&=&\frac{-\sqrt{4-\Gamma ^2} \Gamma  \sin \left(2 \tilde{t}_1\right)+\Gamma ^2 \cos \left(2 \tilde{t}_1\right)-3 \Gamma ^2+8}{2 \sin \left(\tilde{t}_1\right) \Gamma  \sqrt{4-\Gamma ^2}},
\end{array}
\end{equation}
see the appendix for the details of the calculation. But just after the first impulse it is $\phi=0$, thus $\tilde{\lambda}_v(0)=2$, and we finally get
\begin{equation}
\label{cosh(x1)}
\cosh\left(\frac{t_1 \Gamma}{4}\right)=\frac{2 A_1-\sqrt{-A_1^2 B_1^2+B_1^4+4 B_1^2}}{A_1^2-B_1^2}.
\end{equation}
This is a transcendental equation for the switching time $t_1$, where the decay rate $\Gamma$ appears as a parameter.

The corresponding equation for switching time $t_2$ is obtained by propagating forward the adjoint system (\ref{ly_t}), (\ref{lv_t}) in the interval $t_2<t<T$. The initial conditions are $\tilde{\lambda}_v(t_2)=\tilde{\lambda}_v(t_1)=2$ and $\tilde{\lambda}_y(t_2)=\tilde{\lambda}_y(t_1)=\Gamma$, while we use $y(t)$ from Eq. (\ref{y3_opt}) and $y_s=y(t_2)$. We find
\begin{equation}
\label{ly(T)}
\tilde{\lambda}_v(T)=A_2 \cosh\left[\frac{\left(T-t_2\right) \Gamma}{4}\right]+B_2 \sinh\left[\frac{\left(T-t_2\right) \Gamma}{4}\right],
\end{equation}
where
\begin{equation}
\label{A2_B2}
\begin{array}{ccc}
A_2&=&\cos \left(\tilde{T}-\tilde{t}_2\right)-\frac{\Gamma  \sin \left(\tilde{T}-\tilde{t}_2\right)}{\sqrt{4-\Gamma ^2}},\\[7pt]
B_2&=&\frac{\sqrt{4-\Gamma ^2} \Gamma  \sin \left[2 \left(\tilde{T}-\tilde{t}_2\right)\right]+\Gamma ^2 \cos \left[2 \left(\tilde{T}-\tilde{t}_2\right)\right]-3 \Gamma ^2+8}{2 \sin \left(\tilde{T}-\tilde{t}_2\right) \Gamma  \sqrt{4-\Gamma ^2}}.
\end{array}
\end{equation}
From the condition $\phi(T)=0$, equivalent to $\tilde{\lambda}_v(T)=2$, we get a transcendental equation similar to Eq. (\ref{cosh(x1)}),
\begin{equation}
\label{cosh(x2)}
\cosh\left[\frac{\left(T-t_2\right) \Gamma}{4}\right]=\frac{2 A_2-\sqrt{-A_2^2 B_2^2+B_2^4+4 B_2^2}}{A_2^2-B_2^2}.
\end{equation}



\subsection{Smoother controls obtained with reverse engineering}

The previously presented solutions to the spring problem include impulses in the control $u(t)$. Here, we will employ reverse engineering in order to find smoother solutions. Specifically, we will prescribe a sufficiently smooth functional form of $y(t)$, using for example a truncated polynomial series, with enough terms (parameters) to satisfy the boundary conditions (\ref{start}), (\ref{end}) and the integral condition (\ref{impulse}), as well as additional parameters to be used in the minimization of the cost (\ref{cost}). Having determined the optimal parameter values, the corresponding control can be found from Eq. (\ref{osc_equ}) as
\begin{equation}
\label{eq_poly_u}
    u(t)=-\frac{y}{2}-\Gamma  \dot{y}-2 \ddot{y}.
\end{equation}
The initial conditions which we will use are
\begin{equation}
\label{poly_init}
    y(0)=\dot{y}(0)=\ddot{y}(0)=0,
\end{equation}
where the first two are equivalent to conditions (\ref{start}) while the third one implies that $u(0)=0$, contrary to the previous solutions starting with impulses.
The final conditions for our system are
\begin{equation}
\label{poly_final}
    y(T)=\dot{y}(T)=\ddot{y}(T)=0,
\end{equation}
where again the first two are equivalent to conditions (\ref{end}) while the third one implies that $u(T)=0$. The above six boundary conditions are complemented with the pulse area condition (\ref{impulse}), giving seven conditions in total.

We consider $y(t)$ as an $N$-degree polynomial function of dimensionless time $t/T$
\begin{equation}
\label{poly_y}
    y(t)=\sum_{n=0}^{N}a_n \left(\frac{t}{T}\right)^n,
\end{equation}
where $a_n$ are unknown parameters to be determined by the constraints and the optimization procedure. Since there are seven constraints to be satisfied and we additionally need at least one free parameter for the optimization, we conclude that we should use $N\geq 7$. The more free parameters available, the better is the control over the system, in principle. In the following we present the details for the case $N=12$, corresponding to six free parameters, but also give the results for the cases $7\leq N\leq 12$. 





From the initial conditions (\ref{poly_init}) we easily get
\begin{equation}
\label{a0_a1_a2}
    a_0=a_1=a_2=0
\end{equation}
From the final conditions (\ref{poly_final}) and the integral constraint (\ref{impulse}) we obtain, after some manipulation, the relations

\begin{equation}
\label{a3_a4_a5}
\mathbf{a_{3,6}}=\mathbf{A'} \mathbf{a_{7,12}} + \mathbf{c'},
\end{equation}
where
\begin{eqnarray}
\mathbf{a_{3,6}} &=&
\left(\begin{array}{cccc}
a_3 & a_4 & a_5 & a_6
\end{array}\right)^T, \label{dependent}\\
\mathbf{a_{7,12}} &=&
\left(\begin{array}{cccccc}
a_7 & a_8 & a_9 & a_{10} & a_{11} & a_{12}
\end{array}\right)^T, \label{free}
\end{eqnarray}

\begin{equation}
\label{A_2}
\mathbf{A'}=\left(\begin{array}{cccccc}
              \frac{1}{2} & \frac{16}{9} & 4 & \frac{80}{11} & \frac{35}{3} & \frac{224}{13}\\[3pt]
              -\frac{5}{2} & -18 & -50 & -\frac{25}{3} & -\frac{350}{11} & -\frac{945}{13}  \\[3pt]
              \frac{9}{2} & \frac{40}{3} & 27 & \frac{504}{11} & 70 & \frac{1296}{13}       \\[3pt]
              -\frac{7}{2} & -\frac{70}{9} & -14 & -\frac{245}{11} & -\frac{98}{3} & -\frac{588}{13}
         \end{array}\right)
\end{equation}
and
\begin{equation}
\label{c_1}
\mathbf{c'}=
\left(\begin{array}{cccc}
-\frac{140 \pi}{T} & \frac{420 \pi}{T} & -\frac{420 \pi}{T} & \frac{140 \pi}{T}
\end{array}\right)^T
\end{equation}
Constraints (\ref{a0_a1_a2}) and (\ref{a3_a4_a5}) can be expressed in compact form as
\begin{equation}
\label{linear_constraints}
\mathbf{A}^T\mathbf{a}=\mathbf{c},
\end{equation}
where
\begin{equation}
\mathbf{A}^T=
\left(\begin{array}{ccc}
\mathbf{I_{3\times 3}} & \mathbf{O_{3\times 4}} & \mathbf{O_{3\times 6}} \\
\mathbf{O_{4\times 3}} & \mathbf{I_{4\times 4}} & -\mathbf{A'}
\end{array}\right),
\end{equation}
\begin{equation}
\mathbf{a}=
\left(\begin{array}{cccc}
a_0 & a_1 & \ldots & a_{12}
\end{array}\right)^T
\end{equation}
is the vector containing all parameters
and
\begin{equation}
\mathbf{c}=
\left(\begin{array}{cccc}
0 & 0 & 0 & \mathbf{c'}^T
\end{array}\right)^T.
\end{equation}
The symbols $\mathbf{I_{n\times n}}, \mathbf{O_{m\times n}}$ denote a $n\times n$ unit matrix and a $m\times n$ zero matrix, respectively.

Using the polynomial form (\ref{poly_y}), the cost (\ref{cost}) can be expressed as


\begin{table}[t]
\centering
\begin{tabular}{|c || c | c | c|}
 \hline
 $N$ & 8 (2 free) & 10 (4 free) & 12 (6 free) \\
 \hline \hline
 $a_{7}$  & $-\frac{10710 \pi}{T}$ & $-\frac{526680 \pi }{T}$ & $-\frac{9009000 \pi }{T}$ \\ [8pt]
 \hline
 $a_{8}$  & $\frac{5355 \pi}{2 T}$ & $\frac{2317392 \pi }{5 T}$ & $\frac{74666592 \pi }{5 T}$ \\ [8pt]
 \hline
 $a_{9}$  & - & $-\frac{1106028 \pi }{5 T}$ & $-\frac{80258178 \pi }{5 T}$ \\ [8pt]
 \hline
 $a_{10}$ & - & $\frac{1106028 \pi }{25 T}$ & $\frac{270197928 \pi }{25 T}$ \\ [8pt]
 \hline
 $a_{11}$ & - & - & $-\frac{4144140 \pi }{T}$ \\ [8pt]
 \hline
 $a_{12}$ & - & - & $\frac{690690 \pi }{T}$ \\ [8pt]
 \hline\hline
 Cost & $\frac{735\pi^2}{572}\frac{\Gamma}{T}$ & $\frac{6468 \pi ^2}{5525}\frac{\Gamma}{T}$ & $\frac{9009 \pi ^2}{8075}\frac{\Gamma}{T}$ \\ [8pt]
 \hline
  Cost & $1.285\pi^2\frac{\Gamma}{T}$ & $1.171\pi ^2\frac{\Gamma}{T}$ & $1.116\pi ^2\frac{\Gamma}{T}$ \\ [8pt]
 \hline
\end{tabular}
\caption{Optimal free polynomial coefficients that minimize the cost function (\ref{cost_vector_form}), along with the corresponding minimum value, for each of the three cases with 2,4, and 6 free parameters.}
\label{poly_par}
\end{table}

\begin{equation}
\label{cost_vector_form}
J=\Gamma\int_{0}^{T} y^2(t) \,dt=\Gamma T \mathbf{a^T Q a}=\Gamma T\sum_{n=0}^{12}\sum_{m=0}^{12} Q_{nm}a_n a_m,
\end{equation}
where the symmetric matrix $\mathbf{Q}$ has elements

\begin{equation}
\label{Q_nm}
Q_{nm}= \frac{1}{n+m+1} .
\end{equation}

We observe that the minimization of the cost has been brought to the standard form \cite{Boyd04} of minimizing the quadratic form (\ref{cost_vector_form}) under the linear constraints (\ref{linear_constraints}).
This minimization can be easily carried out with respect to the free parameters (\ref{free}), and their corresponding optimal values are displayed in the last column of Table \ref{poly_par}. In the first two columns of this table are also shown the optimal free parameters for the cases $N=10$ and $N=8$. Note that for $N=11$ the optimization procedure gives $a_{11}=0$, and the rest of the free parameters are the same as for the case $N=10$. Similarly, for $N=9$ we find $a_9=0$ and the rest of the free parameters are the same as for $N=8$. Finally, for $N=7$ (the minimum allowed degree), the optimization gives $a_7=0$ while the rest of the parameters are determined from the constraints. Another interesting observation is that the displayed coefficients, as well as the nonzero coefficients $a_3-a_6$ which depend on them through Eq. (\ref{a3_a4_a5}), are independent of the decay rate $\Gamma$. The reason is that, when using Eq. (\ref{eq_poly_u}) in the integral constraint (\ref{impulse}), the term containing $\Gamma$ becomes zero due to the vanishing boundary conditions on $y$. Of course, the corresponding control (\ref{eq_poly_u}) depends on $\Gamma$ through the term $\Gamma\dot{y}$. The minimum cost obtained in each case is displayed in the last two rows of the table, in fractional exact form and approximate decimal form.

From the decimal form of the cost it appears that as $N$ increases the limiting value $\pi^2\Gamma/T$ is approached. We can understand this behavior in the following way. If we integrate both sides of Eq. (\ref{osc_equ}) and use the boundary conditions for $y, \dot{y}$, we get
\begin{equation}
\label{area_constraint}
\int_0^Ty(t)dt=-2\int_0^Tu(t)dt=-\pi.
\end{equation}
Minimizing the cost (\ref{cost}) under constraint (\ref{area_constraint}) corresponds to minimizing the volume of revolution generated by $y(t)$ for a given area (\ref{area_constraint}) below this curve (the negative sign in the area constraint appears because $y<0$). The augmented cost function for this constrained optimization problem is
\begin{equation}
\int_0^TL(y)dt=\int_0^T(\Gamma y^2+\nu y)dt,
\end{equation}
where $\nu$ is the constant Lagrange multiplier corresponding to the integral constraint (\ref{area_constraint}). From the Euler-Lagrange equation we get for the optimal $y^*$
\begin{equation}
\frac{dL}{dy}\rvert_{y = y^*}=0\Rightarrow y^*=\mbox{constant},
\end{equation}
and from the area condition (\ref{area_constraint}) we find $y^*(t)=-\pi/T$, which leads to the cost
\begin{equation}
\label{ultimate}
\Gamma\int_0^Ty^{*2}(t)dt=\pi^2\frac{\Gamma}{T}.
\end{equation}

Of course, in obtaining this cost we have completely ignored the dynamics of the problem, as dictated by the oscillator equation (\ref{osc_equ}) or the equivalent system (\ref{spring_1}), thus this value is rather to be used as a reference. This minimum cost can be achieved for unbounded control $u(t)$ in the following way. A positive impulse is initially applied in order to immediately build a large negative velocity. Then, the system is left to evolve freely for an infinitesimal duration, until position $y=-\pi/T$ is reached, and subsequently a negative impulse is applied to nullify the velocity. The system is held there for the whole duration $T$. Then, a negative impulse is applied, so the system acquires a large positive velocity, followed by an infinitesimal time interval of free evolution which brings the system at $y=0$, and a final positive impulse nullifying the velocity. Note that an essential part of this pulse sequence are the negative impulses, which are forbidden if the constraint (\ref{pos_u}) is on. As we shall see in the next section, the smooth polynomial controls leading to costs approaching the value (\ref{ultimate}), contain intervals where $u(t)<0$, mimicking thus this behavior.

\section{Application to the original spin system}

\label{sec:results}



\begin{table}[t]
\centering
\begin{tabular}{|c || c | c |}
 \hline
 Characteristic & Suboptimal sequence  & Optimal sequence \\
 \hline \hline
 $v_1$ & 0.1914 & 0.2138 \\ [8pt]
 \hline
 $v_2$ & 0.1635 & 0.1036 \\ [8pt]
 \hline
 $v_3$  & - & 0.1108 \\ [8pt]
 \hline
 $v_4$ & - & 0.1842 \\ [8pt]
 \hline
 $u_s$ & 0.0887 & 0.0838 \\ [8pt]
 \hline
 $t_1$ & 3.0454 & 4.1808 \\ [8pt]
 \hline
 $t_2$ & 16.7543 & 15.6159 \\ [8pt]
 \hline
\end{tabular}
\caption{Characteristics (impulses, singular control, normalized switching times) of the intuitive sub-optimal pulse sequence (\ref{almost_optimal_control}) (second column) and the optimal pulse sequence (\ref{optimal_control}) (third column), for decay rate $\Gamma=0.1\Omega_0$ and total duration $T=20/\Omega_0$.}
\label{opti_pulses}
\end{table}

\begin{figure*}[h]
\centering
\begin{subfigure}[b]{.4\textwidth}
    \centering\caption{}\includegraphics[width=\linewidth]{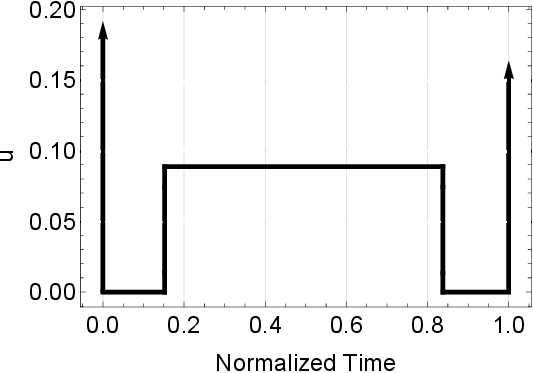}\label{fig:ISO_u}
\end{subfigure}
\hspace{.2cm}
\begin{subfigure}[b]{.4\textwidth}
    \centering\caption{}\includegraphics[width=\linewidth]{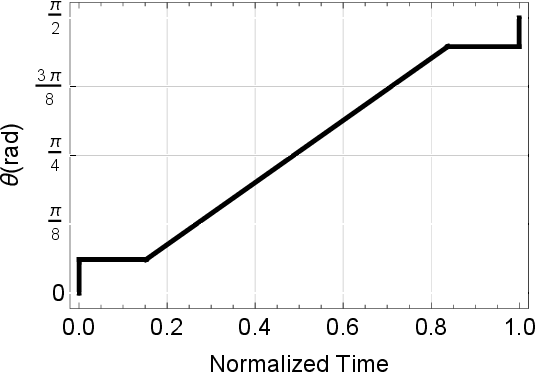}\label{fig:ISO_theta}
\end{subfigure}
\begin{subfigure}[b]{.4\textwidth}
    \centering\caption{}\includegraphics[width=\linewidth]{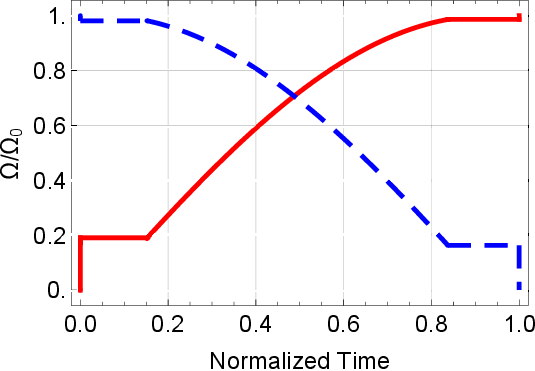}\label{fig:ISO_omegas}
\end{subfigure}
\hspace{.2cm}
\begin{subfigure}[b]{.4\textwidth}
    \centering\caption{}\includegraphics[width=\linewidth]{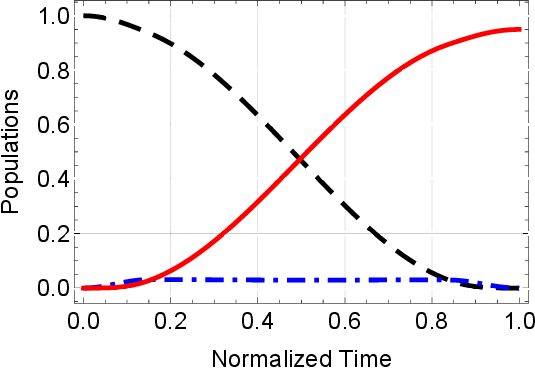}\label{fig:ISO_pop}
\end{subfigure}
\begin{subfigure}[b]{.4\textwidth}
    \centering\caption{}\includegraphics[width=\linewidth]{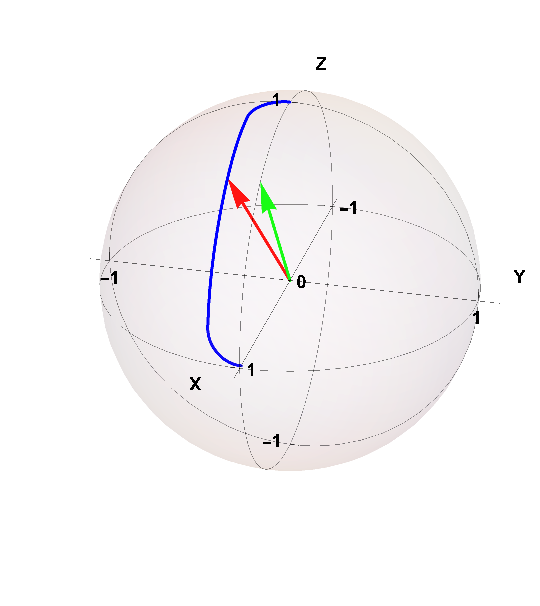}\label{fig:ISO_sphere}
\end{subfigure}
\caption{(a) Intuitive sub-optimal spring control (\ref{almost_optimal_control}), calculated for decay rate $\Gamma/\Omega_0=0.1$ and duration $T=20/\Omega_0$. The arrows indicate control impulses. (b) Evolution of the polar angle $\theta$ of the total field. (c) Corresponding control fields $\Omega_p$ (red solid line), $\Omega_s$ (blue dashed line). (d) Populations of the initial state $\lvert1\rangle$ (black dashed line), intermediate state $\lvert2\rangle$ (blue dashed-dotted line) and target state $\lvert3\rangle$ (red solid line). (e) Trajectory of the spin system in the original reference frame $XYZ$ (blue solid line) along with a visual representation of the Bloch state (red vector) and the field vector (green vector) at time $t=0.4T$.}
\label{fig:ISO}
\end{figure*}

\begin{figure*}[h]
\centering
\begin{subfigure}[b]{.4\textwidth}
    \centering\caption{}\includegraphics[width=\linewidth]{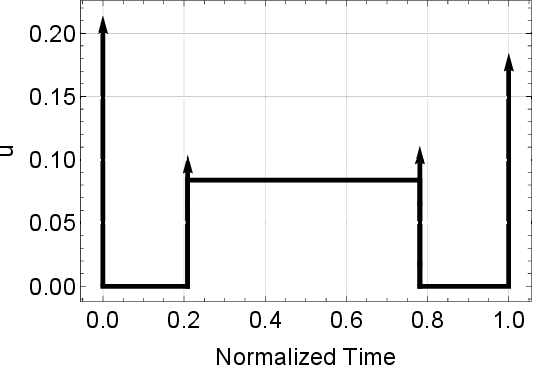}\label{fig:OC_u}
\end{subfigure}
\hspace{.2cm}
\begin{subfigure}[b]{.4\textwidth}
    \centering\caption{}\includegraphics[width=\linewidth]{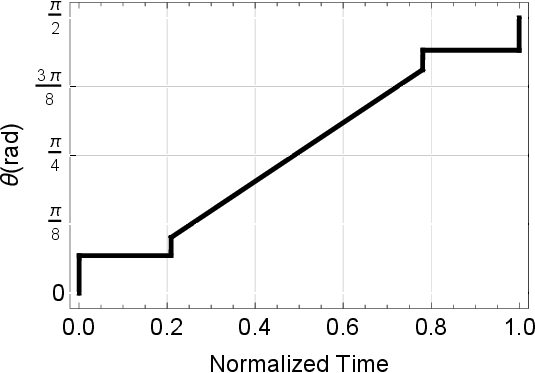}\label{fig:OC_theta}
\end{subfigure}
\begin{subfigure}[b]{.4\textwidth}
    \centering\caption{}\includegraphics[width=\linewidth]{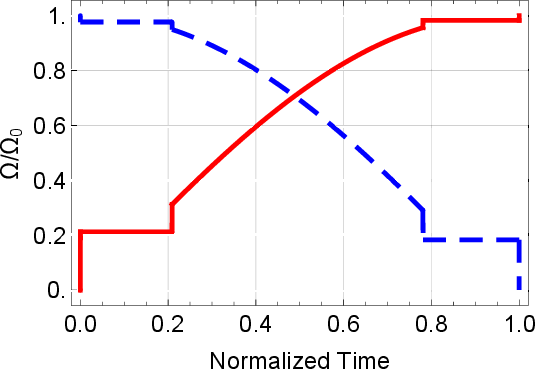}\label{fig:OC_omegas}
\end{subfigure}
\hspace{.2cm}
\begin{subfigure}[b]{.4\textwidth}
    \centering\caption{}\includegraphics[width=\linewidth]{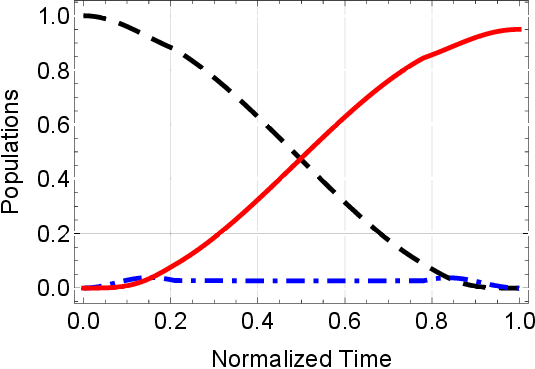}\label{fig:OC_pop}
\end{subfigure}
\begin{subfigure}[b]{.4\textwidth}
    \centering\caption{}\includegraphics[width=\linewidth]{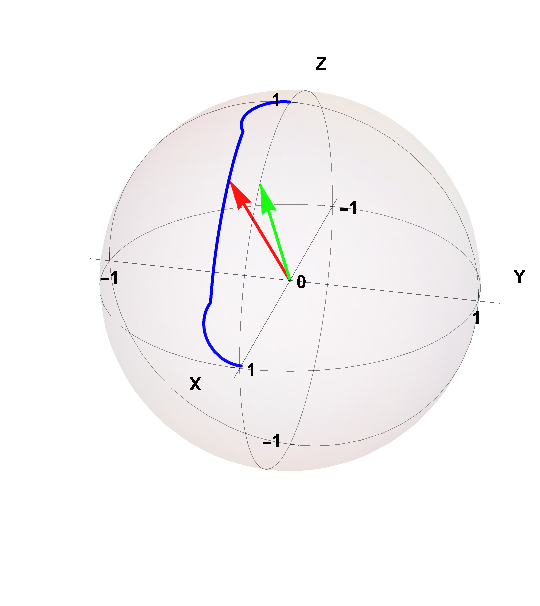}\label{fig:OC_sphere}
\end{subfigure}
\caption{(a) Optimal spring control (\ref{optimal_control}), calculated for decay rate $\Gamma/\Omega_0=0.1$ and duration $T=20/\Omega_0$. The arrows indicate control impulses. (b) Evolution of the polar angle $\theta$ of the total field. (c) Corresponding control fields $\Omega_p$ (red solid line), $\Omega_s$ (blue dashed line). (d) Populations of the initial state $\lvert1\rangle$ (black dashed line), intermediate state $\lvert2\rangle$ (blue dashed-dotted line) and target state $\lvert3\rangle$ (red solid line). (e) Trajectory of the spin system in the original reference frame $XYZ$ (blue solid line) along with a visual representation of the Bloch state (red vector) and the field vector (green vector) at time $t=0.4T$.}
\label{fig:OC}
\end{figure*}

In this section we present specific examples where the controls derived in the previous section using the spring system are applied to the original spin system (\ref{XYZ}).
In all the examples below we use the values $\Gamma=0.1\Omega_0$ and $T=20/\Omega_0$ for the decay rate of state $\lvert 2\rangle$ and the pulse sequence duration, respectively. Using these parameter values in the results of Sec. \ref{sec:idealcase} we find the characteristics (impulses, singular control, switching times) of the intuitive sub-optimal pulse sequence (\ref{almost_optimal_control}) and the optimal pulse sequence (\ref{optimal_control}), which are displayed in Table \ref{opti_pulses}.  In Fig. \ref{fig:ISO_u} we display the intuitive sub-optimal pulse sequence (\ref{almost_optimal_control}), with impulses only at the initial and final times, while in Fig. \ref{fig:ISO_theta} we show the corresponding evolution of the mixing angle $\theta$ and in Fig. \ref{fig:ISO_omegas} the control fields $\Omega_p, \Omega_s$. Observe that the initial and final impulses in $u$ result in finite jumps in the angle $\theta$ and the control fields. These jumps in the controls can be avoided by simply starting and ending the fields to the corresponding nonzero values, thus the intuitive suboptimal pulses are essentially continuous functions. The populations of states $\lvert1\rangle, \lvert2\rangle, \lvert3\rangle$ are depicted in Fig. \ref{fig:ISO_pop}, while the trajectory of the system in the original reference frame $XYZ$ is shown in Fig. \ref{fig:ISO_sphere}, along with a visual representation of the Bloch state vector and the total field vector at time $t=0.4T$. Similar results are displayed in Fig. \ref{fig:OC} for the optimal pulse sequence (\ref{optimal_control}). Comparing the singular control values for the two cases, given in Table \ref{opti_pulses}, we observe that a smaller value corresponds to the optimal case. But recall that the singular control corresponds to the ``force" needed to maintain the oscillator at a constant displacement $y_s$, see Eq. (\ref{us}), thus $|y_s|$ is smaller for the optimal case, a fact that also holds in general for the spin system where $y(t)$ is not exactly constant on the singular arc, leading to a better population efficiency.

\begin{figure*}[h]
 \centering
	\includegraphics[width=0.5\linewidth]{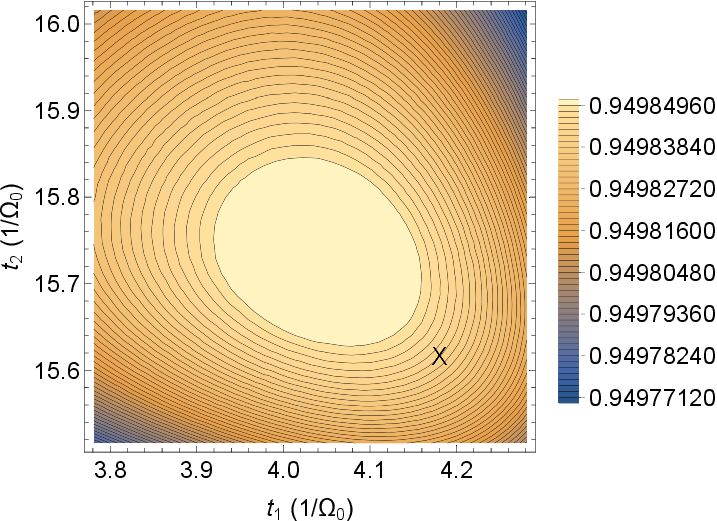}
\caption{Population of state $\lvert3\rangle$ at the final time $T=20/\Omega_0$, in the presence of dissipation $\Gamma/\Omega_0=0.1$, as a function of the switching times $t_1$ and $t_2$, when the impulses and the singular control in pulse sequence (\ref{optimal_control}) are expressed in terms of $t_1, t_2$ using Eqs. (\ref{opt_par_all}) and (\ref{th(T)_opt}). The black X mark denotes the times extracted from the exact solutions of the transcendental Eqs. (\ref{cosh(x1)}), (\ref{cosh(x2)}), which are optimal for the spring system.}
\label{fig:eff_contour}
\end{figure*}

Next, we use Eqs. (\ref{opt_par_all}) and (\ref{th(T)_opt}) obtained for the optimal \emph{spring} case, to express the impulses $v_1, v_2, v_3, v_4$ and the constant singular control $u_s$ in terms of the switching times $t_1,t_2$, and then numerically optimize with respect to these times the population transfer efficiency for the \emph{spin} system. The resultant efficiency is displayed as a contour plot in Fig. \ref{fig:eff_contour}, where the cross mark indicates the values $t_1,t_2$ from the third column of Table \ref{opti_pulses}, which are optimal for the \emph{spring} system. From the colorbar on the side we observe that for the larger portion of the contour plot the difference from the numerically obtained maximum efficiency is only at the fifth decimal point.

\begin{figure*}[h]
\centering
\begin{subfigure}[b]{0.4\textwidth}
    \centering\caption{}\includegraphics[width=\linewidth]{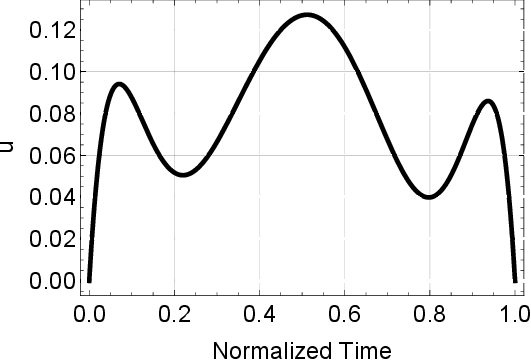}\label{fig:SSO_2_u}
\end{subfigure}
\hspace{.2cm}
\begin{subfigure}[b]{0.4\textwidth}
    \centering\caption{}\includegraphics[width=\linewidth]{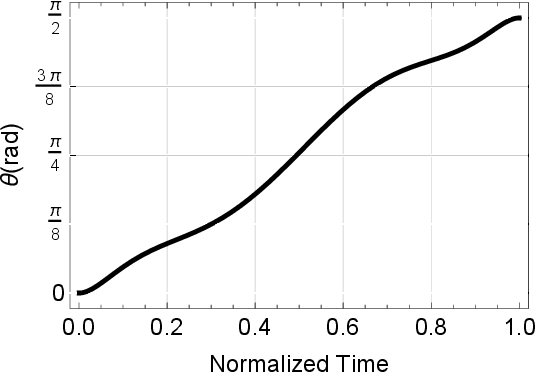}\label{fig:SSO_2_theta}
\end{subfigure}
\begin{subfigure}[b]{0.4\textwidth}
    \centering\caption{}\includegraphics[width=\linewidth]{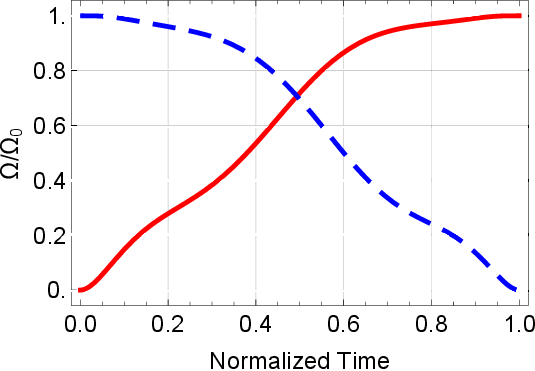}\label{fig:SSO_2_omegas}
\end{subfigure}
\hspace{.2cm}
\begin{subfigure}[b]{0.4\textwidth}
    \centering\caption{}\includegraphics[width=\linewidth]{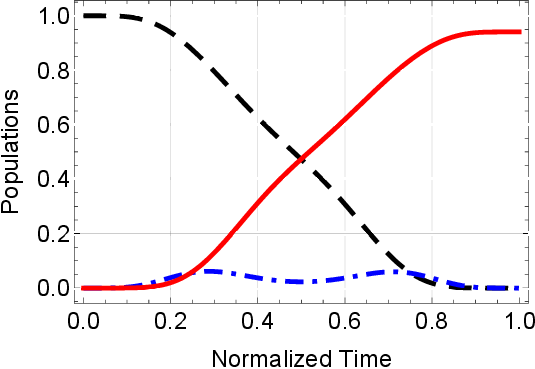}\label{fig:SSO_2_pop}
\end{subfigure}
\begin{subfigure}[b]{0.4\textwidth}
    \centering\caption{}\includegraphics[width=\linewidth]{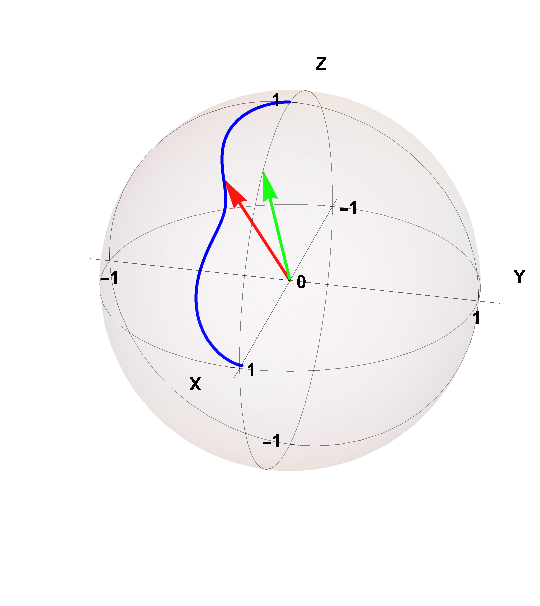}\label{fig:SSO_2_sphere}
\end{subfigure}
\caption{(a) Polynomial control with $N=8$ (two free parameters, see Table \ref{poly_par}), for $T=20/\Omega_0$ and $\Gamma/\Omega_0=0.1$. (b) Evolution of the polar angle $\theta$ of the total field. (c) Corresponding control fields $\Omega_p$ (red solid line), $\Omega_s$ (blue dashed line). (d) Populations of the initial state $\lvert1\rangle$ (black dashed line), intermediate state $\lvert2\rangle$ (blue dashed-dotted line) and target state $\lvert3\rangle$ (red solid line). (e) Trajectory of the system in the original reference frame $XYZ$ (blue solid line) along with a visual representation of the Bloch state (red vector) and the field vector (green vector) at time $t=0.4T$.}
\label{fig:SSO_2}
\end{figure*}

\begin{figure*}[h]
\centering
\begin{subfigure}[b]{0.4\textwidth}
    \centering\caption{}\includegraphics[width=\linewidth]{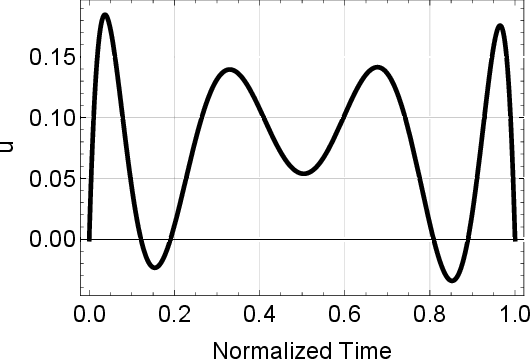}\label{fig:SSO_4_u}
\end{subfigure}
\hspace{.2cm}
\begin{subfigure}[b]{0.4\textwidth}
    \centering\caption{}\includegraphics[width=\linewidth]{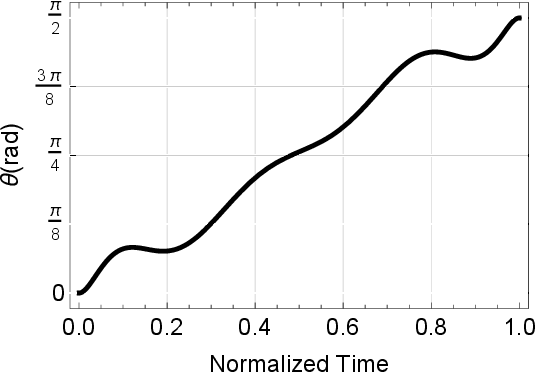}\label{fig:SSO_4_theta}
\end{subfigure}
\begin{subfigure}[b]{0.4\textwidth}
    \centering\caption{}\includegraphics[width=\linewidth]{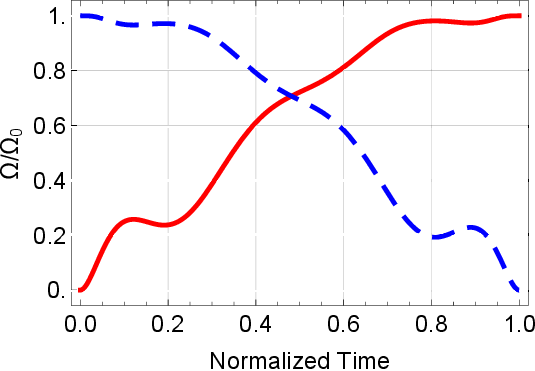}\label{fig:SSO_4_omegas}
\end{subfigure}
\hspace{.2cm}
\begin{subfigure}[b]{0.4\textwidth}
    \centering\caption{}\includegraphics[width=\linewidth]{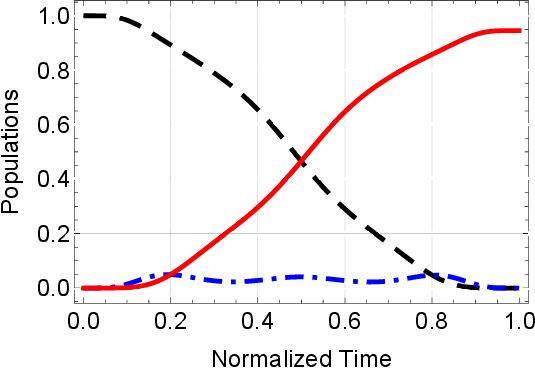}\label{fig:SSO_4_pop}
\end{subfigure}
\begin{subfigure}[b]{0.4\textwidth}
    \centering\caption{}\includegraphics[width=\linewidth]{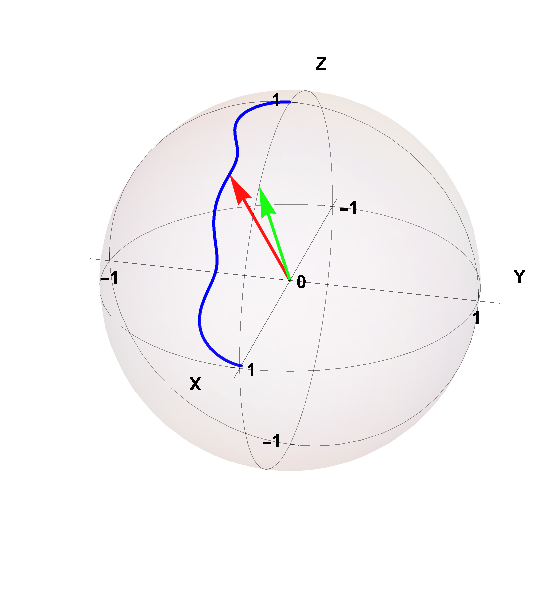}\label{fig:SSO_4_sphere}
\end{subfigure}
\caption{(a) Polynomial control with $N=10$ (four free parameters, see Table \ref{poly_par}), for $T=20/\Omega_0$ and $\Gamma/\Omega_0=0.1$. (b) Evolution of the polar angle $\theta$ of the total field. (c) Corresponding control fields $\Omega_p$ (red solid line), $\Omega_s$ (blue dashed line). (d) Populations of the initial state $\lvert1\rangle$ (black dashed line), intermediate state $\lvert2\rangle$ (blue dashed-dotted line) and target state $\lvert3\rangle$ (red solid line). (e) Trajectory of the system in the original reference frame $XYZ$ (blue solid line) along with a visual representation of the Bloch state (red vector) and the field vector (green vector) at time $t=0.4T$.}
\label{fig:SSO_4}
\end{figure*}

\begin{figure*}[h]
\centering
\begin{subfigure}[b]{0.4\textwidth}
    \centering\caption{}\includegraphics[width=\linewidth]{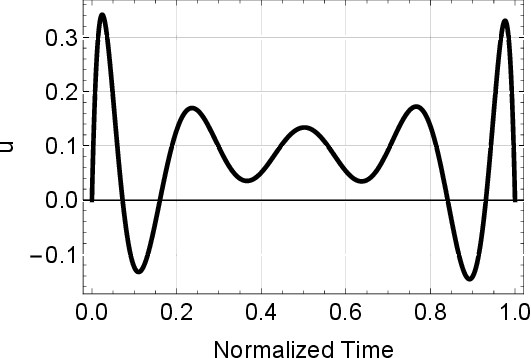}\label{fig:SSO_6_u}
\end{subfigure}
\hspace{.2cm}
\begin{subfigure}[b]{0.4\textwidth}
    \centering\caption{}\includegraphics[width=\linewidth]{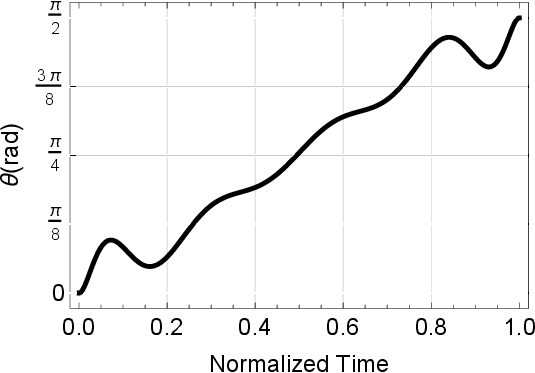}\label{fig:SSO_6_theta}
\end{subfigure}
\begin{subfigure}[b]{0.4\textwidth}
    \centering\caption{}\includegraphics[width=\linewidth]{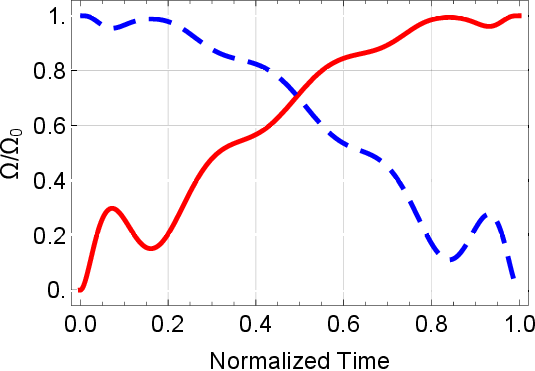}\label{fig:SSO_6_omegas}
\end{subfigure}
\hspace{.2cm}
\begin{subfigure}[b]{0.4\textwidth}
    \centering\caption{}\includegraphics[width=\linewidth]{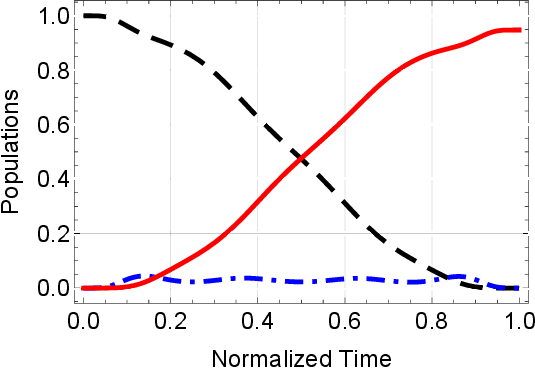}\label{fig:SSO_6_pop}
\end{subfigure}
\begin{subfigure}[b]{0.4\textwidth}
    \centering\caption{}\includegraphics[width=\linewidth]{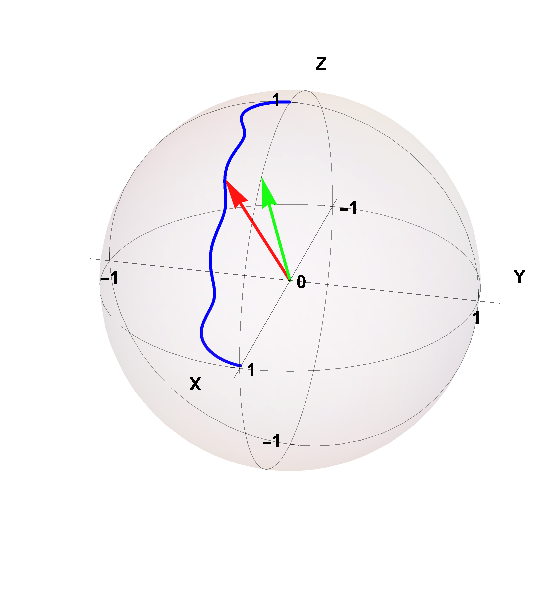}\label{fig:SSO_6_sphere}
\end{subfigure}
\caption{(a) Polynomial control with $N=12$ (six free parameters, see Table \ref{poly_par}), for $T=20/\Omega_0$ and $\Gamma/\Omega_0=0.1$. (b) Evolution of the polar angle $\theta$ of the total field. (c) Corresponding control fields $\Omega_p$ (red solid line), $\Omega_s$ (blue dashed line). (d) Populations of the initial state $\lvert1\rangle$ (black dashed line), intermediate state $\lvert2\rangle$ (blue dashed-dotted line) and target state $\lvert3\rangle$ (red solid line). (e) Trajectory of the system in the original reference frame $XYZ$ (blue solid line) along with a visual representation of the Bloch state (red vector) and the field vector (green vector) at time $t=0.4T$.}
\label{fig:SSO_6}
\end{figure*}

In Figs. \ref{fig:SSO_2}, \ref{fig:SSO_4}, \ref{fig:SSO_6}, we present results analogous to those of Figs. \ref{fig:ISO}, \ref{fig:OC}, but for the polynomial controls with $N=8, 10, 12$, respectively. For $N=8$ we see that the control is positive, Fig. \ref{fig:SSO_2_u}, but for $N=10$ we observe at the beginning and end the development of negative pulses, which follow the initial and precede the final positive pulses, see Fig. \ref{fig:SSO_4_u}. This emerged pulsed sequence is consistent with that described at the end of Sec. \ref{sec:idealcase}. For $N=12$ this behavior becomes even more evident, since the positive and negative pulses become stronger, see Fig. \ref{fig:SSO_6_u}.

\begin{figure*}[h]
 \centering
 \begin{subfigure}[b]{0.4\textwidth}
    \centering\caption{}\includegraphics[width=\linewidth]{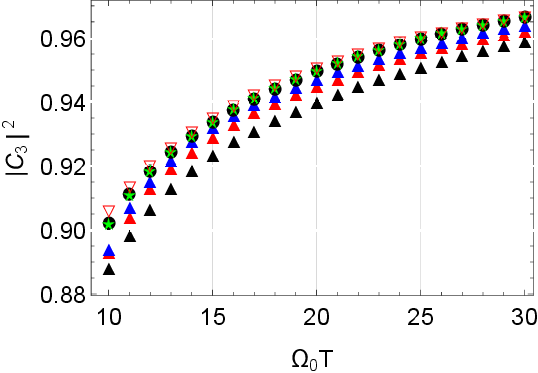}\label{fig:eff_to_T_01}
\end{subfigure}
\hspace{.2cm}
\begin{subfigure}[b]{0.4\textwidth}
    \centering\caption{}\includegraphics[width=\linewidth]{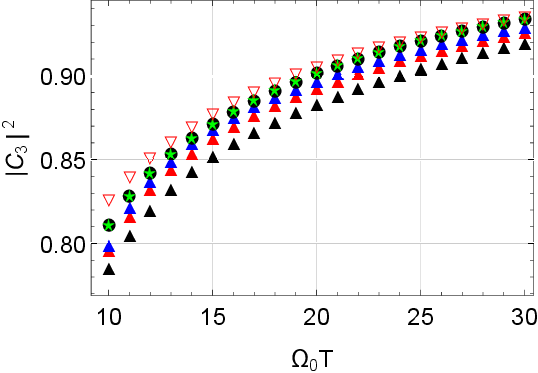}\label{fig:eff_to_T_02}
\end{subfigure}
\caption{Final population of state $\lvert3\rangle$ as a function of the total duration $\Omega_0 T$ in the presence of dissipation (a) $\Gamma/\Omega_0=0.1$ and (b) $\Gamma/\Omega_0=0.2$, for the various control methodologies: numerical optimal spin control \cite{Stefanatos22royal} (hollowed red reversed triangles), optimal spring control (\ref{optimal_control}) (green stars), intuitive suboptimal spring control (\ref{almost_optimal_control}) (black circles), and polynomial spring control with 2, 4, or 6 free parameters (black, red and blue triangles, respectively).}
\label{fig:eff_to_T}
\end{figure*}

Finally, in Fig. \ref{fig:eff_to_T} we plot the population transfer efficiency achieved with the various spring control methods, when applied to the original spin system, versus the duration in the interval $10/\Omega_0\leq T\leq 30/\Omega_0$ with step $\Delta T=1/\Omega_0$, and compare with the efficiency obtained when numerical optimal control \cite{bocop} is directly applied to the spin system (hollowed red reversed triangles) \cite{Stefanatos22royal}. Fig. \ref{fig:eff_to_T_01} is obtained for $\Gamma/\Omega_0=0.1$, while \ref{fig:eff_to_T_02} for $\Gamma/\Omega_0=0.2$. We first observe that in both cases the efficiencies of the optimal spring method (green stars) and the intuitive suboptimal method (black circles) are very close to each other. Among the polynomial controls (triangles of various colors), the one with most free parameters (blue triangles) lies closer in the efficiency. The lower efficiencies obtained with polynomial controls is the price paid for not using impulses, which can quickly build a nonzero population in the intermediate state, facilitating thus the transfer. This inferior performance of polynomial controls relative to the optimal one is consistent with the observations of Ref. \cite{Martikyan20a}, where such controls are applied in a model linear system. Now, in comparison with the numerical optimal spin control (hollowed red reversed triangles), we observe that for both decay rates the optimal spring controls approach its efficiency for longer durations. For the case with smaller decay rate $\Gamma/\Omega_0=0.1$, Fig. \ref{fig:eff_to_T_01}, the approach is accomplished for shorter durations. Of course, the advantage of the spring solutions are the analytical expressions for the impulses, the off control intervals, and the singular control. Note also that, in the case for example of a transmon qubit \cite{Kumar16}, where the states $|1\rangle, |2\rangle, |3\rangle$ are identified as the ground, first and second excited levels, respectively, it is $\Omega_0/2\pi\approx 40$ MHz, thus $\Omega_0\approx 250$ MHz, and $\Gamma\approx 5$ MHz, consequently $\Gamma/\Omega_0=0.02$, much smaller than the values used here. We close by pointing out that for all the presented methodologies the population transfer efficiency increases with duration, as can be observed from Fig. \ref{fig:eff_to_T}. The reason is that for all these strategies the population transfer for increased duration is accomplished by less populating the lossy intermediate state.

\section{Conclusion}

\label{sec:conclusion}

We derived shortcuts to adiabaticity which maximize population transfer in a three-level $\Lambda$ quantum system, using the spin to spring mapping to formulate the corresponding optimal control problem on the simpler system of a classical driven dissipative harmonic oscillator. We solved the spring optimal control problem and obtained analytical expressions for the impulses, the durations of the zero control intervals and the singular control, which compose the optimal pulse sequence. We also derived suboptimal solutions for the spring problem, one with less impulses than the optimal and others with smoother polynomial controls. We applied the solutions derived for the spring system to the original system, and compared the population transfer efficiency with that obtained for the original system using numerical optimal control. For all dissipation rates used, the efficiency of the optimal spring control approached that of the numerical optimal solution for longer durations, with the approach accomplished earlier for smaller decay rates. The efficiency achieved with the suboptimal spring control with less impulses was very close to that of the optimal spring control in all cases, while that obtained with polynomial controls lied below, and this is the price paid for not using impulses, which can quickly build a nonzero population in the intermediate state. The analysis of the optimal solution for the classical driven dissipative oscillator is not restricted to the system at hand but can also be applied in the transport of a coherent state trapped in a moving harmonic potential and the transport of a mesoscopic object in stochastic thermodynamics. Note that there are numerous works in the literature dealing with the problem of ``optimizing" STIRAP, many of which we cite in the Introduction. In each of them different constraints are used and different optimization goals are set. Sometimes, even different systems are employed. For example, in Ref. \cite{Vasilev09}, unitary evolution (without dissipation) is considered and STIRAP pulses are optimized with respect to a measure of nonadiabaticity. In other shortcut works, a direct coupling between states $|1\rangle$ and $|3\rangle$ is introduced to cancel the diabatic transitions. From these observation becomes obvious that a fair comparison between the various methodologies would be a very difficult, if not impossible, task. That said, we believe that the approach followed here and in Ref. \cite{Stefanatos22royal}, to minimize losses or directly maximize the population of the target state in the presence of dissipation, appears to be the most straightforward. And definitely these methods are optimal under the assumed constraints and for the system used, given in Eq. (\ref{Schrodinger}) or the equivalent Eq. (\ref{XYZ}).

\appendix*

\section{Calculation of $\lambda_v(0)$, $\lambda_v(T)$}

We substitute $y(t)$ from Eq. (\ref{y1_opt}) into Eq. (\ref{ly_t}) and use the backward time $\sigma=t_1-t$. The initial conditions for the backward evolution are $\tilde{\lambda}_v(\sigma=0)=\tilde{\lambda}_v(t=t_1)=2$ and $\tilde{\lambda}_y(\sigma=0)=\tilde{\lambda}_y(t=t_1)=\Gamma$. We then apply the Laplace transform to system (\ref{ly_t}), (\ref{lv_t}) in the backward time, taking into account the corresponding initial conditions, and obtain


\begin{equation}
\label{YV_Lap_Eqs1}
\begin{array}{ccc}
-s \Lambda_y(s)&=&-\Gamma + \frac{\Lambda_v(s)}{4} +\frac{\Gamma -4 s+\sqrt{4-\Gamma ^2} \cot \left(\tilde{t}_1\right)}{8 s^2-4 \Gamma  s+2},\\[5pt]
-s \Lambda_v(s)&=&\frac{\Gamma  \Lambda_v(s)}{2}-\Lambda_y(s)-2,
\end{array}
\end{equation}
where note that the minus signs in the left hand side appear because we consider the backward evolution.
We then solve for $\Lambda_y(s)$ and $\Lambda_v(s)$ finding


\begin{equation}
\label{YV_Lap_Sol1}
\begin{array}{ccc}
\Lambda_y(s)&=&\frac{16\Gamma s^3 -2\Gamma\left(2\Gamma^2-5\right)s +\Gamma^2 -2 -(\Gamma +2 s)\sqrt{4-\Gamma ^2} \cot \left(\tilde{t}_1\right)}{16 s^4 -4 \left(\Gamma ^2-2\right) s^2 +1},\\[5pt]
\Lambda_v(s)&=&\frac{32 s^3 -8\left(\Gamma^2-2\right) s +2\Gamma -2\sqrt{4-\Gamma ^2} \cot \left(\tilde{t}_1\right)}{16 s^4 -4 \left(\Gamma ^2-2\right) s^2 +1}.
\end{array}
\end{equation}
Taking the inverse Laplace transform of $\Lambda_v(s)$ and evaluating it at $\sigma=t_1$, corresponding to $t=0$, we find the expression for $\tilde{\lambda}_v(t=0)$ given in Eq. (\ref{lv(0)}).

Following a similar process in the interval $t_2\leq t\leq T$, we substitute $y(t)$ from Eq. (\ref{y3_opt}) into Eq. (\ref{ly_t}) and then apply the Laplace transform to system (\ref{ly_t}), (\ref{lv_t}), now in the forward time, taking into account the boundary conditions $\tilde{\lambda}_v(t_2)=2$ and $\tilde{\lambda}_y(t_2)=\Gamma$. We obtain


\begin{equation}
\label{YV_Lap_Eqs2}
\begin{array}{ccc}
s \Lambda_y(s)&=&\Gamma + \frac{\Lambda_v(s)}{4} -\frac{\Gamma +4 s-\sqrt{4-\Gamma ^2} \cot \left(\tilde{T}-\tilde{t}_2\right)}{8 s^2+4 \Gamma  s+2},\\[5pt]
s \Lambda_v(s)&=&\frac{\Gamma  \Lambda_v(s)}{2}-\Lambda_y(s)+2,
\end{array}
\end{equation}
and after solving for $\Lambda_y(s)$ and $\Lambda_v(s)$ we find


\begin{equation}
\label{YV_Lap_Sol2}
\begin{array}{ccc}
\Lambda_y(s)&=&\frac{16 \Gamma s^3 -16\Gamma^2 s^2 +2\Gamma \left(2\Gamma^2+3\right)s -3\Gamma^2 -\sqrt{4-\Gamma ^2} (\Gamma -2 s) \cot \left(\tilde{t}_1\right)+2}{\left(4 s^2-2 \Gamma  s+1\right)^2},\\[5pt]
\Lambda_v(s)&=&\frac{32 s^3 -32\Gamma s^2 +8\left(\Gamma^2+2\right) s -6\Gamma -2 \sqrt{4-\Gamma ^2} \cot \left(\tilde{t}_1\right)}{\left(4 s^2-2 \Gamma  s+1\right)^2}.
\end{array}
\end{equation}
Taking the inverse Laplace transform of $\Lambda_v(s)$ and evaluating it at $t=T$, we find the expression (\ref{ly(T)}) for $\lambda_v(T)$.

\bibliographystyle{apsrev4-2}
\bibliography{bibliography}

\vspace*{1. cm}

\end{document}